\documentclass[prb,twocolumn]{revtex4-1}

\usepackage{amssymb}
\usepackage{amsmath}
\usepackage{graphicx}

\begin{document}

\title{Scaling relations for magnetic nanoparticles}
\author{P. Landeros, J. Escrig and D. Altbir}
\affiliation{Departamento de F\'{\i}sica, Universidad de Santiago de Chile, USACH, Av.
Ecuador 3493, Santiago, Chile}
\author{D. Laroze}
\affiliation{Instituto de F\'{\i}sica, Pontificia Universidad Cat\'olica de
Valpara\'{\i}so, Casilla 4059, Valpara\'{\i}so, Chile}
\author{J. d'Albuquerque e Castro}
\affiliation{Instituto de F\'{\i}sica, Universidade Federal do Rio de Janeiro, Cx.Postal
68.528, 21941-972, RJ, Brazil.}
\author{P. Vargas}
\affiliation{Departamento de F\'{\i}sica, Universidad T\'{e}cnica Federico
Santa Mar\'{\i}a, Casilla 110-V, Valpara\'{\i}so, Chile}

\begin{abstract}
A detailed investigation of the scaling relations recently proposed by
d'Albuquerque e Castro \textit{et al}.\cite{prl} to study the magnetic
properties of nanoparticles is presented. Analytical expressions for the
total energy of three characteristic internal configurations of the
particles are obtained, in terms of which the behavior of the magnetic phase
diagram for those particles upon scaling of the exchange interaction is
discussed. The exponent $\eta $ in scaling relations is shown to be
dependent on the geometry of the vortex core, and results for specific cases
are presented.
\end{abstract}

\maketitle

\section{Introduction}

In recent years, a great deal of attention has been focused on the study of
regular arrays of magnetic particles produced by nano-imprint lithography.
Besides the basic scientific interest in the magnetic properties of these
systems, there is evidence that they might be used in the production of new
magnetic devices, or as media for high density magnetic recording \cite{Chou}%
. One of the main points in the study of such systems concerns the internal
magnetic structure of the nanoparticles as a function of their shape and
size. For example, in the case of cylindrically shaped particles produced by
electrodeposition, the internal arrangements of the magnetic moments have
been identified as been close to one of the following three (idealized)
characteristic configurations, namely ferromagnetic with the magnetization
parallel to the basis of the cylinder ($F1$), ferromagnetic with the
magnetization parallel to the cylinder axis ($F2$), and a vortex state, in
which most of the magnetic moments lie parallel to the basis of the cylinder
($V$) \cite{Chapman, Ross0}. The occurrence of each of these configurations
depends on geometrical factors, such as the linear dimensions of the
cylinders and their aspect ratio. Clearly, for the development of magnetic
devices based on those arrays, knowledge of the internal magnetic structure
of the particles is of fundamental importance.

Experimentally, attempts have been made to determine, from the analysis of
hysteresis curves\cite{Cowburn,Lebib}, the range of values of diameter $D$
and height $H$ of cylindrically shaped particles for which the internal
arrangement of the magnetic moments is close to either one of the two
ferromagnetic configurations ($F1$ or $F2$) or \ to the vortex one ($V$).
However, such approach does not allow a clear description of the magnetic
structure of individual cylinders, since in many cases the internal magnetic
configurations are not readily identifiable from magnetization curves.

On the other hand, theoretical determination of the configuration of lowest
energy of particles in the size range of those currently produced, based on
a microscopic approach and using present standard computational facilities,
is out of reach. The reason is the exceedingly large number of magnetic
moments within such particles, which may exceed 10$^{9}.$ Recently,
d'Albuquerque e Castro \textit{et al.}\cite{prl} have proposed a scaling
technique for determining the phase diagram giving the configuration of
lowest energy among the three above mentioned characteristic magnetic
configurations. They have shown that such diagram can be obtained from those
for much smaller particles, in which the exchange interaction $J$ has been
scaled down by a factor $x<1$, i.e. for $J^{\prime }=xJ$. The diagram for
the full strength of the exchange interaction is then obtained by scaling up
the $D^{\prime }$ and $H^{\prime }$ axes in the phase diagram for $J^{\prime
}$ by a factor $1/x^{\eta }$. In their work, the exponent $\eta $ has been
determined numerically from the position, as a function of $x$, of a triple
point $(D_{t},H_{t})$ in the phase diagram where the three configurations
have equal energy. The scaling technique has been applied to the
determination of the phase diagram of cylindrically shaped \cite{prl} and
truncated conical\cite{APL} particles. In both cases, $\eta $ turned out to
be approximately equal to $0.55$.

We recall that the vortex configuration exhibits a core region within which
the magnetic moments have a non-zero component parallel to the axis of
either the cylinder or the truncated cone. We remark that the determination
of the geometry of the core (i.e. its shape and size), on the basis of a
microscopic model in which the individual magnetic moments are considered,
would require a prohibitively large computational effort. For this reason,
d'Albuquerque e Castro \textit{et al.}\cite{prl} and Escrig \textit{et al.}%
\cite{APL} adopted a simplified representation of the vortex core,
consisting of a single line of magnetic moments along the axis of either the
cylinders or the truncated cones. The phase diagrams thus obtained are in
good agreement with experimental data, provided appropriate values of the
exchange are considered.

The scaling technique represents a useful tool for studying the magnetic
properties of nanosized particles. It is conceptually simple and rather
interesting from the theoretical point of view. Its implementation depends
on the determination of the exponent $\eta $ in the scaling factor, which so
far has been done numerically. The agreement, within error bars, between the
values of $\eta $ for cylinders and truncated conical particles suggests
that this parameter does not depend on the shape of the particles. However,
there still remains the question regarding the possible dependence of $\eta $
on the geometry of the vortex core. The present work aims precisely at
clarifying this point.

We focus on cylindrically shaped particles, for which a large amount of
experimental data is available. We adopt a continuous model for the internal
magnetic structure of the particles, on the basis of which analytical
results for the total energy in each configuration can be obtained. We use
these results to investigate the behavior of the phase diagrams under
scaling transformation, from which the value of $\eta $ can be determined.
We find that the value of $\eta $ does depend on the geometry of the vortex
core. This point is discussed at length below.

\section{Continuous magnetization model}

We adopt a simplified description of the system, in which the discrete
distribution of magnetic moments is replaced with a continuous one, defined
by a function $\overrightarrow{M}(\vec{r})$ such that $\overrightarrow{M}(%
\vec
{r})\ \delta V$ gives the total magnetization within the element of
volume $\delta V$ centered at $\vec{r}$. This model provides a fairly good
basis for the discussion of the magnetic properties of nanosized particles.
For cylindrically shaped particles, the magnetization density $%
\overrightarrow {M}(\vec{r})$ in the two ferromagnetic configurations, $F1$
and $F2,$ is given by $M_{0}\hat{x}$ and $M_{0}\hat{z}$, respectively. Here $%
M_{0}$ \ is the saturation magnetization density, and $\hat{x}$ and $\hat{z}$
are unit vectors parallel to the basis and to the axis of the cylinders,
respectively. For the vortex configuration, we assume that the magnetization
density has the form%
\begin{equation}
\overrightarrow{M}(\vec{r})=M_{z}(\rho)\ \hat{z}+M_{\varphi}(\rho )\ \hat{%
\varphi}\ ,  \label{mv}
\end{equation}
where $\hat{z}$ and $\hat{\varphi}$ are unit vectors in cylindrical
coordinates, and $M_{z}$ and $M_{\varphi}$ satisfy the relation $%
M_{z}^{2}+M_{\varphi}^{2}=M_{0}^{2}$. Thus, the profile of the vortex core
is fully specified by just giving the function $M_{z}(\rho)$. It is worth
pointing out that the functional form in Eq.(\ref{mv}) does not take into
account the possibility of a dependence of the core shape on coordinate $z$.

We then look at the total energy of the three configurations under
consideration, from which the magnetic phase diagram can be obtained and its
behavior under scaling investigated. We restrict our discussion to arrays in
which the separation between cylinders is sufficiently large for the
interaction between them to be ignored.\cite{Grimsditch,Guslienko}

The internal energy per unit of volume, $E_{tot},$ of a single cylinder is
given by the sum\ of three terms corresponding to the magnetostatic ($%
E_{dip} $), the exchange ($E_{ex}$), and the anisotropy ($E_{K}$)
contributions. However, in the case of particles produced by
electrodeposition, the crystalline anisotropy term is much smaller than the
other two\cite{Ross2}, so its inclusion has little effect on the phase
diagram. In view of that, it will be neglected in our calculations.

\subsection{Ferromagnetic configurations}

Since the exchange term depends only on the relative orientation of the
magnetic moments, it has the same value $E_{ex}^{(F)}$ in the two
ferromagnetic configurations. Since it also appears as an additive term in
the expression for exchange energy in the vortex configuration, it can be
simply left out in our calculations.

The magnetostatic term is generally given by\cite{aharoni} 
\begin{equation}
E_{dip}=\ \frac{\mu _{0}}{2V}\ \int \vec{M}(\vec{r})\cdot \left( \vec{\nabla}%
U\right) \ dV,  \label{edip}
\end{equation}%
where $U(\overrightarrow{r})$ is the magnetostatic potential. In the above
expression, an additive term independent of the configuration has been left
out. For the ferromagnetic configurations we find that%
\begin{equation}
E_{dip}^{(\alpha )}=\frac{1}{2}N_{\alpha }\mu _{0}M_{0}^{2},  \label{edipf}
\end{equation}%
where $\alpha =F1$, $F2$, and $N_{\alpha }$ are the demagnetizing factors,
given in SI unities by\cite{factors} 
\begin{equation}
N_{F1}=\frac{1}{2}\ .\ _{2}F\ _{1}\left[ -\frac{1}{2}\ ,\frac{1}{2}\
,2,-\left( \frac{D}{H}\right) ^{2}\right] -\frac{2D}{3\pi H},  \label{nf1}
\end{equation}%
and 
\begin{equation}
N_{F2}=1-\ _{2}F\ _{1}\left[ -\frac{1}{2}\ ,\frac{1}{2}\ ,2,-\left( \frac{D}{%
H}\right) ^{2}\right] +\frac{4D}{3\pi H}.  \label{nf2}
\end{equation}%
In the above two equations, $_{2}F_{1}(a,b,c,d)$ is a hypergeometric
function. Notice that demagnetizing factors depend on just the ratio $D/H$.

\subsection{Vortex configuration}

Assuming that $\vec{M}(\vec{r})$ varies slowly on the scale of the lattice
parameter, the exchange term for this configuration can be approximated by%
\cite{aharoni}%
\begin{equation*}
E_{ex}^{(V)}=\frac{A}{V}\int \left( \left( \overrightarrow{\nabla }%
m_{x}\right) ^{2}+\left( \overrightarrow{\nabla }m_{y}\right) ^{2}+\left( 
\overrightarrow{\nabla }m_{z}\right) ^{2}\right) \ dV\ ,
\end{equation*}%
where $A$ is the exchange stiffness constant, and $m_{i}=M_{i}/M_{0}$, for $%
i=x,\ y,\ z$. We recall that $A$ is proportional to the exchange interaction
energy $J$ between the magnetic moments.\cite{aharoni} Making use of the
expression for $\overrightarrow{M}(\overrightarrow{r})$ in Eq.(\ref{mv}), we
find 
\begin{equation}
E_{ex}^{(V)}=\ \frac{2A}{R^{2}}\int\limits_{0}^{R}f(\rho )\ \rho \ d\rho
\,\,,  \label{eexv}
\end{equation}%
where $R=D/2$, and $f(\rho )=(\partial m_{z}/\partial \rho
)^{2}/(1-m_{z}^{2})+(1-m_{z}^{2})/\rho ^{2}$, with $m_{z}(\rho )=M_{z}(\rho
)/M_{0}$. The additive term $E_{ex}^{(F)}$ on the r.h.s. of the above
equation has been omitted.

The magnetostatic term can be also written in terms of $\vec{M}(\vec{r})$.
In the vortex configuration, the magnetostatic potential is given by%
\begin{equation*}
U(\vec{r})=\frac{1}{4\pi }\int_{S_{1}}\frac{M_{z}\ (\rho _{1})}{\left\vert 
\vec{r}-\vec{r}_{1}\right\vert }dS_{1}-\frac{1}{4\pi }\int_{S_{2}}\frac{%
M_{z}\ (\rho _{2})}{\left\vert \vec{r}-\vec{r}_{2}\right\vert }dS_{2},
\end{equation*}%
where $S_{1}$ and $S_{2}$ are the surfaces of the top and bottom basis of
the cylinder, respectively. After some manipulations, the expression for $U(%
\vec{r})$\ reduces to%
\begin{multline*}
U(\rho ,z)=\frac{1}{2}\int\limits_{0}^{R}\rho ^{\prime }d\rho ^{\prime
}M_{z}(\rho ^{\prime }) \\
\int\limits_{0}^{\infty }dk\ J_{0}(k\rho )\ J_{0}(k\rho ^{\prime })(-\
e^{-kz}+e^{-k(H-z)}),
\end{multline*}%
where $J_{0}(x)$ is the cylindrical Bessel function of order zero. Taking
this result into Eq.(\ref{edip}), we find\cite{hs}%
\begin{equation}
E_{dip}^{(V)}=\frac{\pi \mu _{0}}{V}\int\limits_{0}^{\infty }dk\left(
\int\limits_{0}^{R}\rho \ J_{0}(k\rho )\ M_{z}(\rho )\ d\rho \right)
^{2}\left( 1-e^{-Hk}\right) .  \label{edipv}
\end{equation}

\section{Total energy calculation and scaling transformation}

At this point, it is necessary to specify the function $M_{z}(\rho )$. Since
no rigorous result regarding the shape of the vortex core is available, we
resort to a simple but physically plausible approximation, given by%
\begin{equation}
M_{z}(\rho )=\left\{ 
\begin{array}{c}
M_{0}\left( 1-(\rho /\rho _{c})^{2}\right) ^{n},\text{ \ \ \ for \ \ }\rho
\leq \rho _{c} \\ 
0\text{ \ \ \ \ \ \ \ \ \ \ \ \ \ \ \ \ \ \ \ \ \ \ \ \ \ otherwise}%
\end{array}%
\right. ,  \label{mz}
\end{equation}%
where $\rho _{c}\leq R$ and $n$ is a non-negative constant. Alternative
expressions for $M_{z}(\rho )$ have been proposed in the literature.\cite{UP}

The above functional form for $M_{z}(\rho )$ allows us to evaluate the
energy integrals in Eqs. (\ref{eexv}) and (\ref{edipv}) analytically. Then,
for integer values of $n$, the expression for $E_{ex}^{(V)}$ in Eq.(\ref%
{eexv}) reduces to

\begin{equation}
E_{ex}^{(V)}=\frac{2A}{R^{2}}\left( \ln \frac{R}{\rho _{c}}+\gamma
_{n}\right) ,  \label{eexvm}
\end{equation}%
where $\gamma _{n}=\frac{1}{2}H\left[ 2n\right] -nH\left[ -\frac{1}{2n}%
\right] $. Here, $H\left[ k\right] =\sum_{i=1}^{k}1/i$ are the harmonic
numbers. For the dipolar energy term in Eq.(\ref{edipv}) we obtain%
\begin{equation}
E_{dip}^{(V)}=\frac{6W_{d}^{0}\rho _{c}^{3}}{HR^{2}}\left( \alpha _{n}-\frac{%
\rho _{c}}{4H}\beta _{n}\text{ }F(n,\frac{\rho _{c}}{H})\right) ,
\label{edipvm}
\end{equation}%
where%
\begin{equation}
\alpha _{n}=\frac{2^{2n-1}\Gamma (n+1)^{3}}{\Gamma (\frac{3}{2}+n)\Gamma (%
\frac{5}{2}+2n)}  \label{alpha}
\end{equation}%
\begin{equation}
\beta _{n}=1/(1+n)^{2}  \label{betha}
\end{equation}%
\begin{equation}
W_{d}^{0}=\frac{1}{6}\mu _{0}M_{0}^{2}  \label{w0}
\end{equation}%
\begin{equation*}
F(n,\frac{\rho _{c}}{H})=\text{\ }_{3}F\,_{2}\left[ \left\{ \frac{1}{2},1,%
\frac{3}{2}+n\right\} ,\left\{ n+2,2n+3\right\} ,-\frac{4\rho _{c}^{2}}{H^{2}%
}\right] \ .
\end{equation*}%
Here, $_{3}F_{2}$ denotes the generalized hypergeometric function.

\section{Results}

Having evaluated all relevant contributions to the total energy in the three
cases of interest, we are in a position to investigate the magnetic phase
diagram for cylinders. In particular, we can look at the position of the
triple point $(D_{t},H_{t})$ as a function of the factor $x$ which scales
the stiffness constant $A$ (or exchange interaction $J$). We notice that
since the energy of the two ferromagnetic configurations, $E_{tot}^{(F1)}$
and $E_{tot}^{(F2)}$, are equal at the triple point, we immediately get the
equation 
\begin{equation*}
N_{F1}(\xi _{t})=N_{F2}(\xi _{t})\ ,
\end{equation*}%
whose solution is $\xi _{t}=D_{t}/H_{t}=1.10317...$ (independent of $A$ or $%
J $).\cite{factors} As a consequence, $D_{t}$ and $H_{t}$ are proportional
and must exhibit the same functional dependence on $x$ (or equivalently, on $%
A$).

We proceed in our analysis by looking at the case considered by
d'Albuquerque e Castro \textit{et al.,}\cite{prl} in which the core radius
is independent of $x$, and of the order of the lattice spacing (first core
model). This corresponds to taking the limit $\rho _{c}\ll R_{t}=D_{t}/2$ in
the expressions for the total energy. In this limit, $\ln \left( R/\rho
_{c}\right) $ becomes much larger in modulus than $\gamma _{n}$, so that the
latter can be safely neglected in Eq.(\ref{eexvm}). Then, Eqs. (\ref{edipf}%
), (\ref{eexvm}), and (\ref{edipvm}) give the following equation for $R_{t}$ 
\begin{equation}
\frac{1}{2}N_{\alpha }\mu _{0}M_{0}^{2}=\frac{2A}{R_{t}^{2}}\ln \frac{R_{t}}{%
\rho _{c}}\ ,  \label{tp1}
\end{equation}%
where $\alpha $ is either $F1$ or $F2$. Now, if we scale down the exchange
interaction by a factor $x<1$, that is to say, if we consider a reduced
exchange stiffness $A^{\prime }=xA$, and assume that $R_{t}$ and the new
radius at the triple point $R_{t}^{\prime }$ are related according to $%
R_{t}^{\prime }=x^{\eta }R_{t},$ we find 
\begin{equation*}
\frac{2A}{R_{t}^{2}}\ \ln \frac{R_{t}}{\rho _{c}}\ =x^{1-2\eta }\ \frac{2A}{%
R_{t}^{2}}\ \ln \frac{x^{\eta }\ R_{t}}{\rho _{c}}\ .
\end{equation*}%
This expression gives us the following equation for $\eta $%
\begin{equation}
\ln \frac{R_{t}}{\rho _{c}}=\frac{\eta }{x^{2\eta -1}-1}\ln x\ .
\label{etha}
\end{equation}

It is clear from this equation that $\eta $ must in all cases be greater
than 0.5. It approaches this lower bound only when $R_{t}$ is much larger
than the lattice spacing (\textit{i.e. }$R_{t}\gg \rho _{c})$, in other
words, when the particles have macroscopic sizes.

\begin{figure}[h]
\includegraphics[width=8cm]{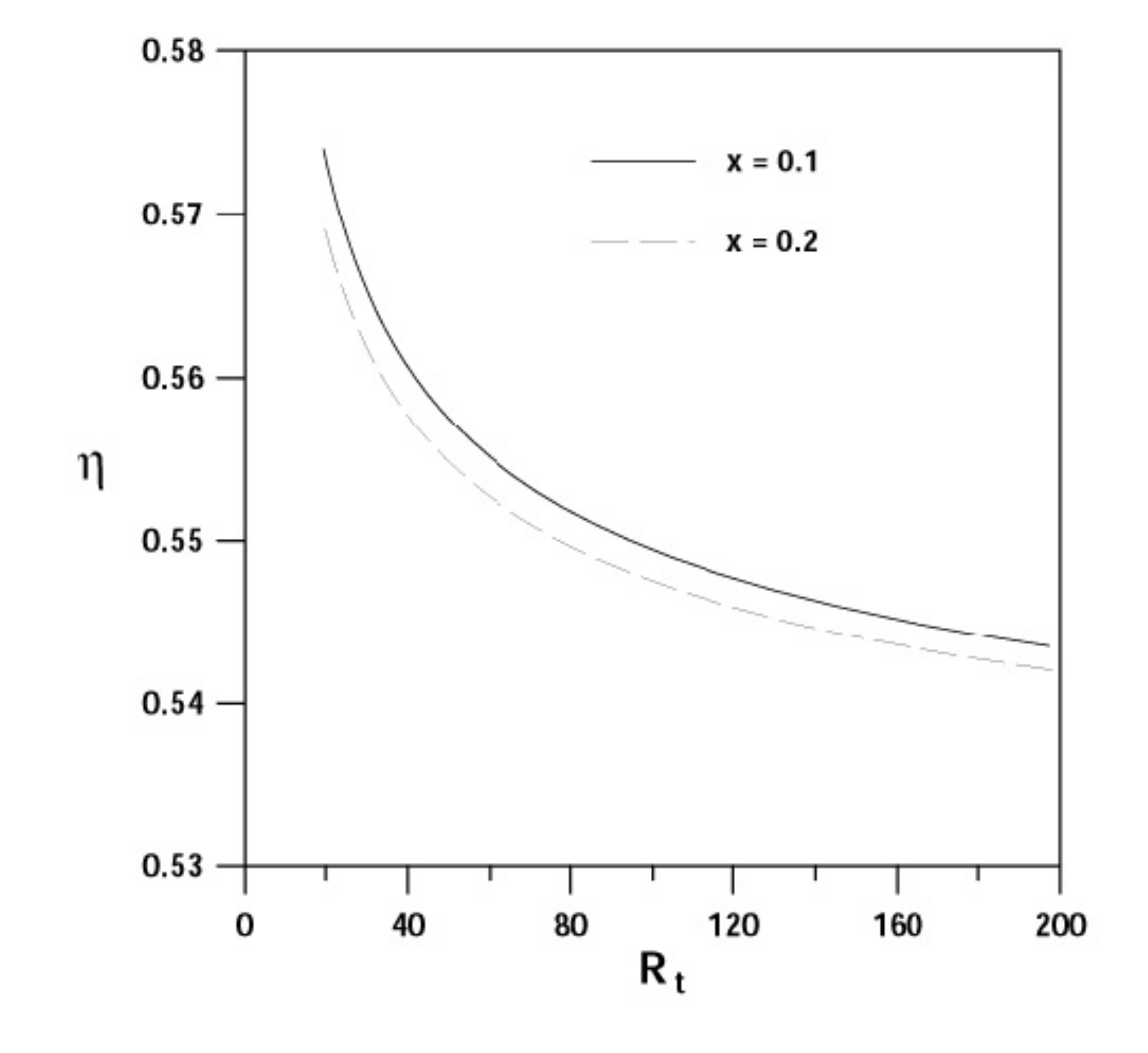}\label{fig1a} %
\includegraphics[width=8cm]{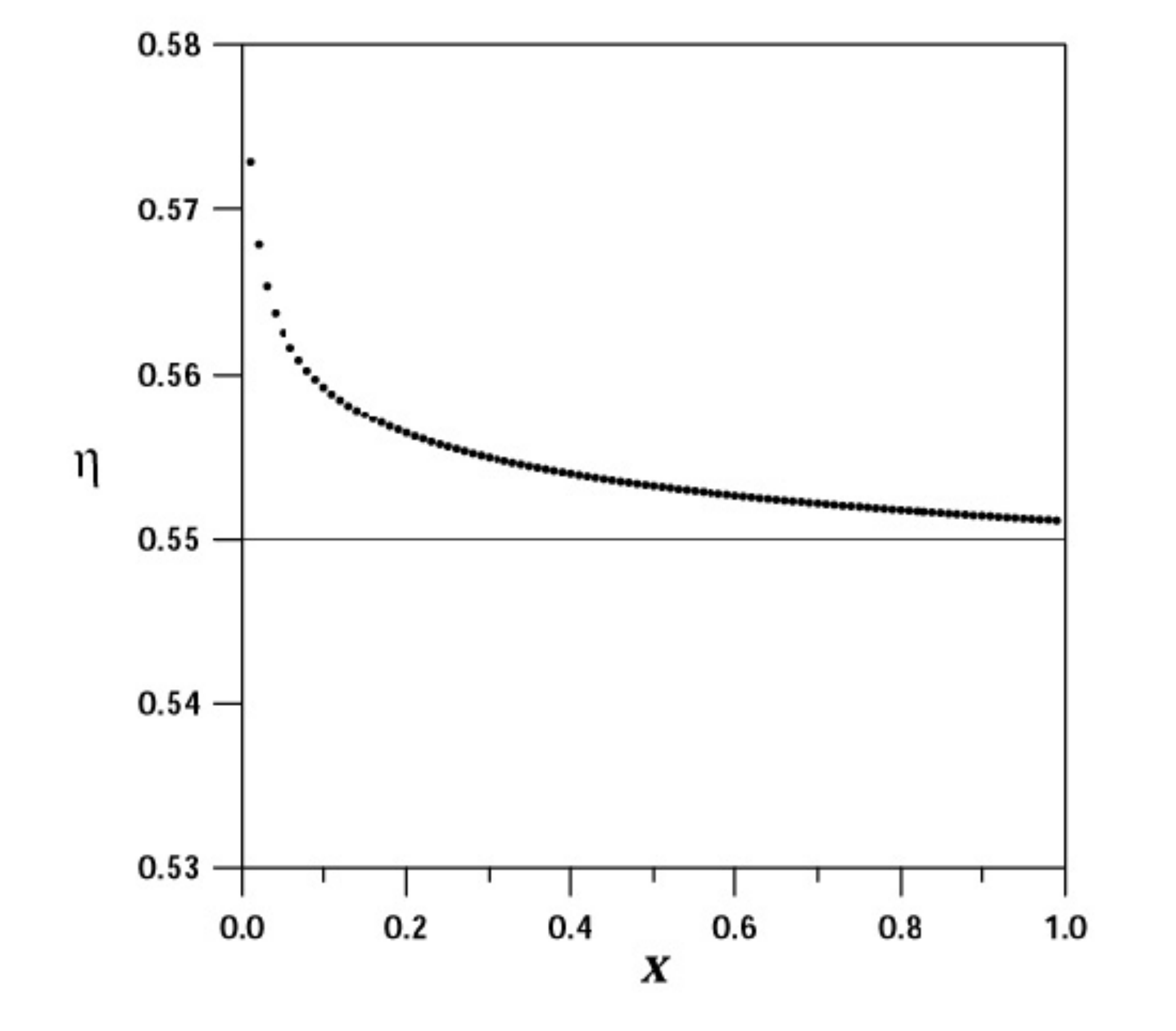}\label{fig1b}
\caption{Exponent $\protect\eta $ given by Eq.(\protect\ref{etha}) plotted
as a function of $R_{t}$, for $x$=0.1 (solid line) and 0.2 (dashed line)
(a), and as a function of $x,$ for $R_{t}$=44 nm (b). }
\end{figure}

The behavior of $\eta $ in Eq.(\ref{etha}) is presented in Fig.(1). Fig.
(1a) shows $\eta $ as a function of $R_{t}$, for 20 nm $\leq R_{t}\leq $ 100
nm, and $\rho _{c}=0.2$ nm. We notice that in this range of $R_{t}$, $%
0.54<\eta <0.58$. It is also interesting to look at the behavior of $\eta $
as a function of $x$. Fig. (1b) shows $\eta $ as a function of $x$, for $%
0.01\leq x\leq 1$, $R_{t}=44$ nm, and $\rho _{c}=$ $0.2$ nm. From the curves
in Figs. (1a) and (1b), we find that for $x\geqslant 0.05$, $\eta $ turns
out to be close to 0.55, as numerically obtained by d'Albuquerque e Castro 
\textit{et al.}\cite{prl}

It is worth commenting on the effect of using a single value of $\eta $, say
0.55, to scale phase diagrams for the core model considered just above. As
already pointed out, the diagram for the full strength of the exchange
interaction can be obtained from the one corresponding to a reduced
interaction $J^{\prime }=xJ$ (with $x<1$) by multiplying the axes $H^{\prime
}$ and $D^{\prime }$ of the latter by $1/x^{\eta }$. Thus, an inaccuracy $%
\delta \eta $ in the value of $\eta $ results in inaccuracies $\delta H$ and 
$\delta D$ in the coordinates in the scaled diagram. Indeed, if we write $%
\eta =\eta _{0}\pm \delta \eta $, with $\delta \eta $/$\eta _{0}\ll 1$, we
immediately get%
\begin{equation*}
\left\vert \frac{\delta H}{H_{0}}\right\vert =-\ \left( \eta _{0}\ln
x\right) \ \left\vert \frac{\delta \eta }{\eta _{0}}\right\vert \ ,
\end{equation*}%
where $H_{0}=x^{\eta _{0}}H^{\prime }$. Since $\eta _{0}\approx 0.55$ and $%
\delta \eta /\eta _{0}\approx 0.01$ (estimated from Fig.(1a)), we find that,
even for $x$ as small as 0.05, the relative error $\delta H/H_{0}$ is
smaller than 2\ \%. Thus, we do not expect large discrepancies between the
calculated phase diagram and the experimental data resulting from such
inaccuracy in $\eta $ since a relative error of 2\% should not exceed the
experimental error.

We remark that the above results for $\eta $ hold also when the core radius
corresponds to several interatomic distances and is kept fixed as the
exchange interaction is scaled up or down.

We next consider the case in which $\rho _{c}$ is adjusted so as to minimize
the energy of the vortex configuration (second core model). From \ Eqs.\text{%
(\ref{eexvm}) }and \text{(\ref{edipvm}) we obtain the following equation for 
}$\rho _{c}$%
\begin{multline*}
3\alpha _{n}\frac{\rho _{c}^{3}}{H^{3}}-\beta _{n}\frac{\rho _{c}^{4}}{H^{4}}%
F(n,\rho _{c}/H) \\
+\frac{\beta _{n}}{2(n+2)}\frac{\rho _{c}^{6}}{H^{6}}G(n,\rho _{c}/H)=\frac{%
2A}{\mu _{0}M_{0}^{2}H^{2}},
\end{multline*}%
where 
\begin{multline*}
G(n,\rho _{c}/H)= \\
\,_{3}F_{2}\left[ \left\{ \frac{3}{2},2,\frac{5}{2}+n\right\} ,\left\{
3+n,4+2n\right\} ,-\frac{4\rho _{c}^{2}}{H^{2}}\right] \ .
\end{multline*}

Eq. (\ref{rhoc}) can be solved numerically for $\rho _{c}$ in terms of $H$, $%
A$, and $n$. We remark that for the core model under consideration, $\rho
_{c}$ does not depend on the radius $R$. This follows from the fact that the
outer region of the cylinder does not interact with the core (apart from the
exchange interaction across the interface between the two regions). As a
consequence, for a given value of $\rho _{c}$, the difference between the
total energy of two cylinders of the same height but different radii does
not depend on $\rho _{c}\,$, hence it does not contribute to the derivative
of $E_{tot}^{(V)}$ with respect to $\rho _{c}$. That is to say, the equation
for $\rho _{c}$ which minimizes the total energy of the vortex configuration
is independent of $R$.

Figure (2) illustrates $M_{z}(\rho )$ for $\ A=87.39$ meV/nm, $%
M_{0}=1.4\times 10^{6}$ A/m, and two values of $H$, namely 20 and 100 nm.
For each $H$, results are presented for $n=$ 2 (dotted line), 4 (dashed
line), and 10 (solid line). The values of $A$ and $M_{0}$ correspond to
those for Co, and have been taken from Ref.[14]. The value of $\rho _{c}$ in
each case has been obtained from Eq.(\ref{rhoc}).

\begin{figure}[h]
\includegraphics[width=8cm]{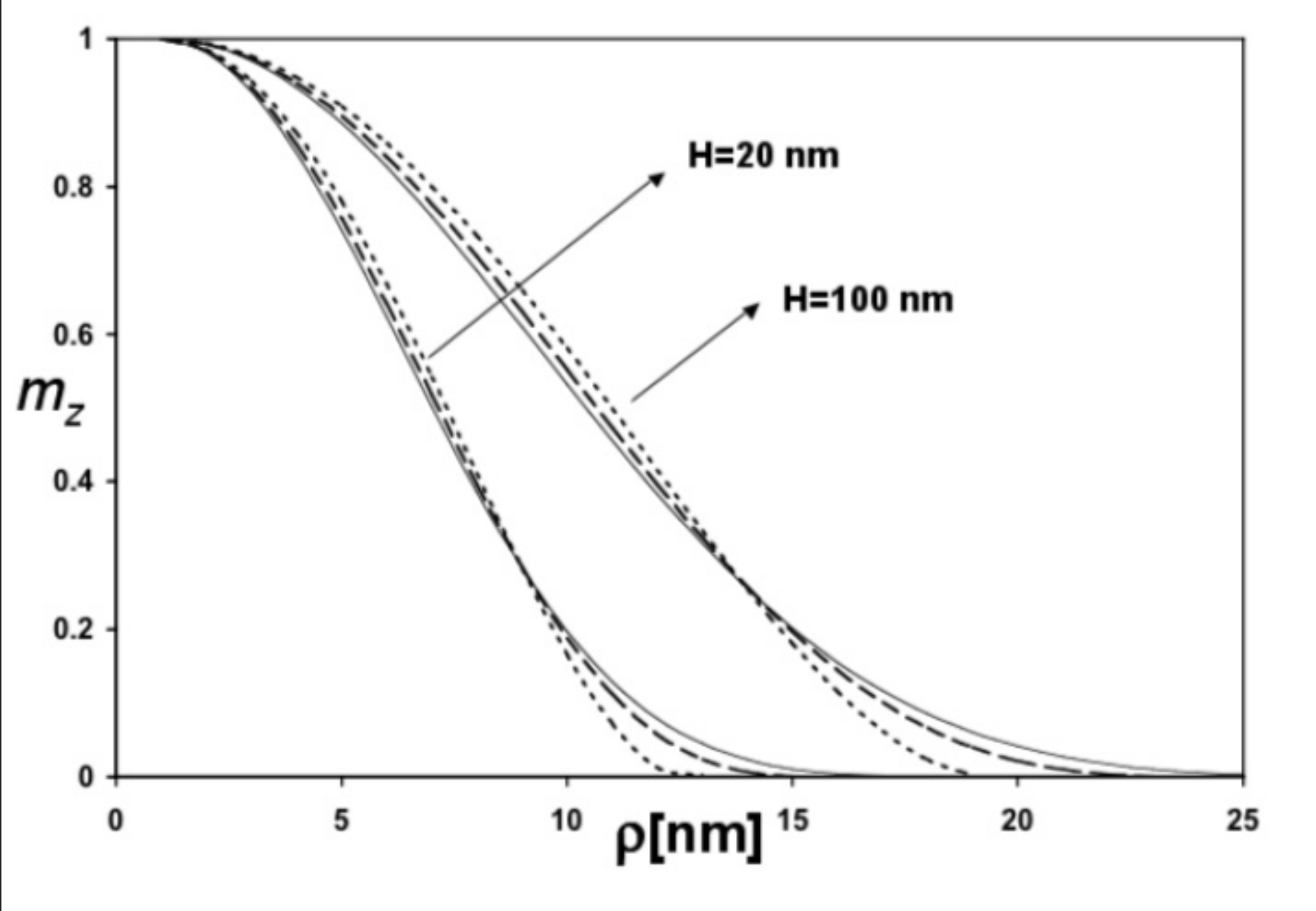}\label{fig2}
\caption{Reduced magnetization $m_{z}=M_{z}/M_{0}$ as a function of $\protect%
\rho $, for $n=2$ (dotted line), 4 (dashed line), and 10 (solid line). The
two sets of curves correspond to $H=$ 20 nm, and 100 nm. Values of $A$ and $%
M_{0}$ have been taken from Ref. [14], and correspond to those for Co.}
\end{figure}

In order to investigate the behavior of \ magnetic phase diagram upon
scaling of the exchange interaction for this second core model, we take $n=$
4, which according to Fig. (2) provides a physically sound description of
the core profile, and calculate the phase diagrams for distinct values of $x$%
. Fig. (3) shows results for cylinders of Co ($A=87.39$ meV/nm and $%
M_{0}=1.4\times 10^{6}$ A/m)\ corresponding to $x=0.12$ (dashed lines) and $%
x=24$ (dotted lines).

\begin{figure}[h]
\includegraphics[width=8cm]{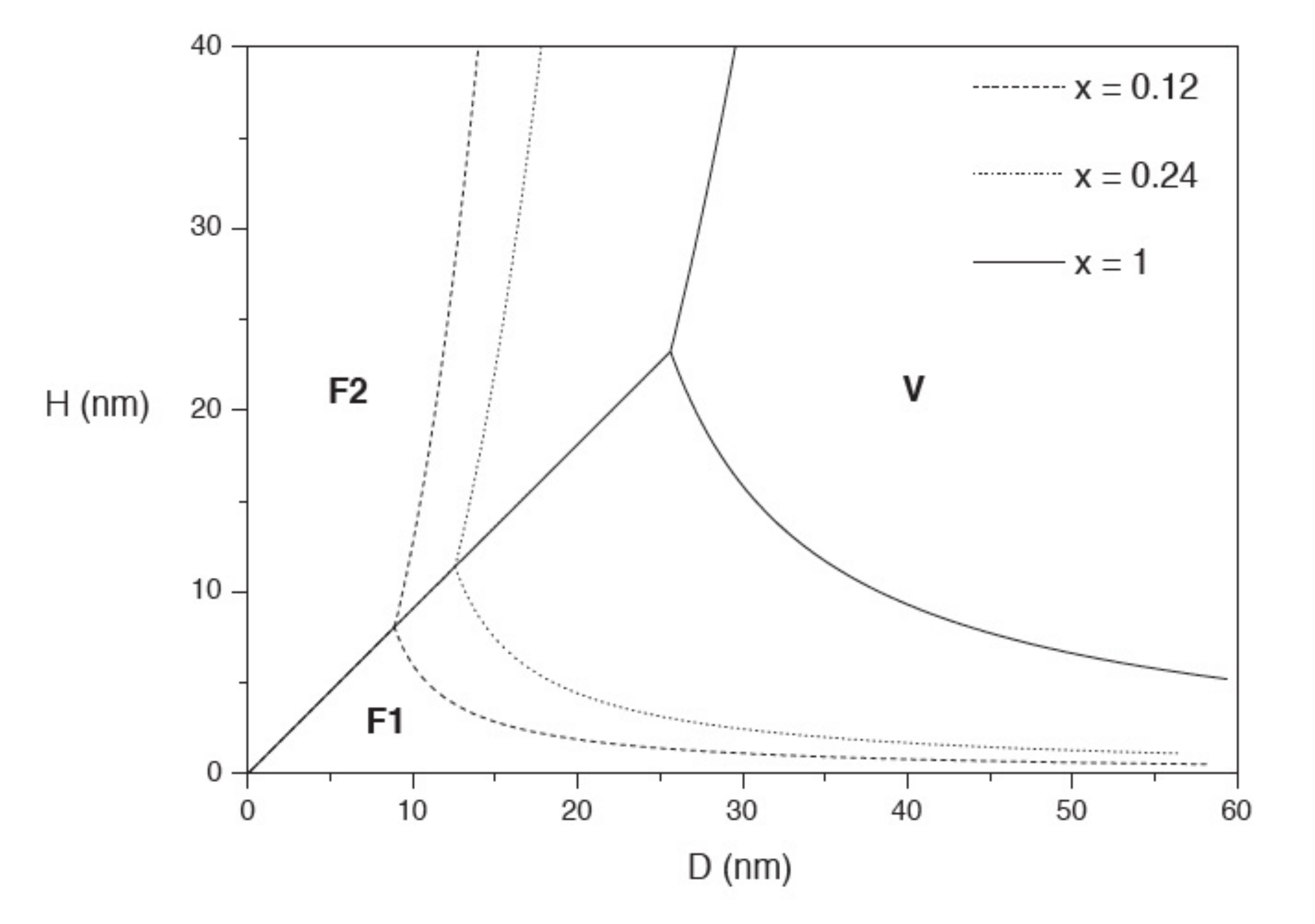}\label{fig3}
\caption{Phase diagram for Co cylinders corresponding to $x=0.12$ (dashed
lines), $x=0.24$ (dotted lines) and $x=1$ (solid lines) obtained for core
model 2 (see text). }
\end{figure}

We then find that, for the present core model, the coordinates $(D_{t},H_{t})
$ of the triple point follow the relations 
\begin{eqnarray}
D_{t}(x) &=&25.61\ x^{0.5} \\
H_{t}(x) &=&23.22\ x^{0.5}
\end{eqnarray}%
in which $\eta =0.5$. The diagram for the the full strength of the exchange
interaction, $x=1$, is represented by solid line.

We remark that these results holds for any other integer values of $n$, the
reason being the fact that since $\rho _{c}$ is adjusted so as to minimize
the energy in the vortex configuration, the effective radius of the core
turns out to be independent of $n$, as clearly shown in Fig.(2).

\section{Conclusions}

We have carried out a detailed analysis of scaling technique recently
proposed by d'Albuquerque e Castro \textit{et al}\cite{prl} to investigate
the magnetic phase diagram of nanoparticles. As already pointed out, this
technique enables us to obtain the phase diagram for particles in the
nanometer size range from those corresponding to much smaller particles, in
which the exchange interaction has been reduced. The scaling technique is
easily implemented and represents a rather useful tool for dealing with
nanoparticle systems. In addition, the existence of scaling relations and
their connection with the model adopted to describe the magnetic particles
bring about interesting theoretical considerations.

The present work sheds light on a very interesting feature of the scaling
relations, namely the dependence of the exponent $\eta $ on the model
adopted for describing the core of the vortex configuration. Based on a
continuous magnetization model, we were able to derive analytical
expressions for the total energy in each configuration, which allowed us to
determine the exponent $\eta $. We found that in the case of nanoparticles
for which the core dimensions, and consequently its contribution to the
total energy, can be either neglected or do not change much upon scaling of $%
A$, $\eta $ turns out to be weakly dependent on $x$ and quite close to 0.55.
Nevertheless, when the contribution from the core is relevant and its size
upon scaling of $A$ changes so as to minimize the total energy in the vortex
configuration, then $\eta $ becomes exactly equal to 0.5.

\section*{Acknowledgments}

This work has been partially supported by Fondo Nacional de Investigaciones
Cient\'{\i}ficas y Tecnol\'{o}gicas (FONDECYT, Chile) under Grants
Nos.1040354, 1020071, 1010127 and 7010127, and Millennium Science Nucleus
"Condensed Matter Physics" P02-054F of Chile, and CNPq, FAPERJ, and
Instituto de Nanoci\^{e}ncias/MCT of Brazil. CONICYT Ph.D. Program
Fellowships, as well as MECESUP USA0108 and FSM-990 projects are gratefully
acknowledged.

\end{document}